\documentclass{article}

\usepackage{arxiv}

\usepackage[utf8]{inputenc} 
\usepackage[T1]{fontenc}    
\usepackage{hyperref}       
\usepackage{url}            
\usepackage{booktabs}       
\usepackage{amsfonts, amsmath}       
\usepackage{nicefrac}       
\usepackage{microtype}      
\usepackage{graphicx}
\usepackage{lipsum}
\usepackage{comment}
\usepackage{color}
\usepackage{bm}
\usepackage{float}
\usepackage{fontawesome}
\usepackage{xspace}
\usepackage{tabularx}
\usepackage{chngcntr}

\usepackage{subcaption}
\usepackage{caption}

\newcommand{\glib}{\texttt{RNAglib}\xspace}
\newcommand{\forgi}{\texttt{forgi}\xspace}

\graphicspath{{figs/}}

\title{RNAglib: A Python package for RNA 2.5D graphs}

\author {

  Vincent Mallet \thanks{Equal contribution} \\
  Pasteur Institute \\
  Les Mines-Paristech \\
  \texttt{vincent.mallet96@gmail.com}
  
  \And
  
  Carlos Oliver$^\ast$ \\
  School of Computer Science, McGill Univerity\\
  Montreal Institute for Learning Algorithms (MILA)\\
  \texttt{carlos.gonzalezoliver@mail.mcgill.ca}
  
  \And
  
  Jonathan Boadbent $^\ast$ \\
  School of Computer Science, McGill University\\
  \texttt{jonathan.broadbent@mail.mcgill.ca}
  
  \And
  
  William L. Hamilton \\
  School of Computer Science, McGill Univerity\\
  Montreal Institute for Learning Algorithms (MILA)\\
  \texttt{wlh@cs.mcgill.ca}
  
  \And
  
  J\'er\^ome Waldisp\"uhl \thanks{To whom correspondence should be addressed.}\\
  School of Computer Science, McGill University\\
  \texttt{jeomew@cs.mcgill.ca}
  }
  \date{} 

\begin{document}
\maketitle

\noindent\hspace{0.15\linewidth}\begin{minipage}{0.7\textwidth}

\section*{Abstract}

RNA 3D architectures are stabilized by sophisticated networks of (non-canonical) base pair interactions, which can be conveniently encoded as multi-relational graphs and efficiently exploited by graph theoretical approaches and recent progresses in machine learning techniques.
\glib is a library that eases the use of this representation, by providing clean data, methods to load it in machine learning pipelines and graph-based deep learning models suited for this representation.
\glib also offers other utilities to model RNA with 2.5D graphs, such as drawing tools, comparison functions or baseline performances on RNA applications.\\
The method and data is distributed as a fully documented pip package. 
A full description for installation and usage is available as a README.\\

\textbf{Availability:} http://rnaglib.cs.mcgill.ca \\
\end{minipage}

\section{Introduction}

Recent developments in machine learning and deep learning techniques enable us to leverage the vast amount of biological data, such as sequencing data or biochemical assays, publicly released and organized by the community. This allowed breakthrough in many areas, including the prediction of protein 3D structures with AlphaFold \cite{jumper2021highly}.

These progresses allow the whole structural biology community to move on new critical challenges, previously out of reach and far from the spotlight.
This is in particular the case for RNAs.
RNA is a highly structured molecule that supports many regulatory and enzymatic functions beyond its well-known messenger role \cite{mattick2006non, fire1998potent}. 
As such, it is a promising class of therapeutic drug targets \cite{yu2020rna, crooke2018rna} as illustrated by novel treatments of CMV retinitus patients with AIDS \cite{vitravene2002randomized} or the production of self-amplifying vaccines, which has recently seen a high-level of success in clinical trials for COVID-19 \cite{fuller2020amplifying}.
Because of the limited (but growing) amount of structural data available for RNAs, the task of designing robust machine learning methods to predict RNA 3D structures is more challenging that for proteins.
However, RNA folding rely on a remarkable hierarchical organization of its structure. 
From our capacity to efficiently use this information will depend the success of machine learning applications.
Representing objects as graphs is a strong prior knowledge and graph neural networks have shown to induce a tremendous performance boost in many applications.

To capture the tertiary structure of RNA in a computationally feasible manner, a growing number of algorithms make use of 2.5D graph networks \cite{bayespairing, oliver2020vernal, islam2021rnamotifcontrast, sarver2008fr3d, reinharz2018mining, djelloul2008automated}.
These networks represent RNA molecules as topological graphs, whose nodes are nucleotides and whose edge types are structural categories of interactions.
We have previously successfully combined the 2.5D graph representation with graph neural networks to predict small molecule binding \cite{oliver2020augmented} and believe that their wider adoption is limited by the lack of dedicated software to use this representation.
To our knowledge, the Python package \forgi  is the only effort in streamlining research on RNA networks.
However, it focuses on coarse-grained models based on secondary structure elements instead of base-pair interactions and does not include machine learning features.

\section{Contribution}

We present a PyPi package, \glib that aims to fill that gap by providing utilities to represent the structure of RNA as 2.5D graphs.
\glib provides clean data available for download along with loading, encoding and splitting routines. 
It also provides structural comparison functions that help unsupervised pre-training, and default models to learn RNA properties such as small-molecules or protein binding : we offer a benchmark performance for these tasks.
Finally, \glib offers utility scripts to save, preprocess or plot graphs so that the manipulation of the data for research is facilitated.

The package includes the following modules:
\begin{itemize}
    \item \texttt{prepare\_data}: Updates and builds annotated graph database.
    \item \texttt{loading}: DGL data loaders for RNA graphs.
    \item \texttt{models}: Pre-built GCN models.
    \item \texttt{learning}: Learning routines for the easiest use of the package.
    \item \texttt{drawing}: 2.5D graph visualization tools.
    \item \texttt{kernels}: Subgraph comparison routines.
    \item \texttt{benchmark}: Reproducible evaluation procedures for benchmarking
\end{itemize}

\section{Data processing}
\begin{figure}
    \centering
    \includegraphics[width=.7\linewidth]{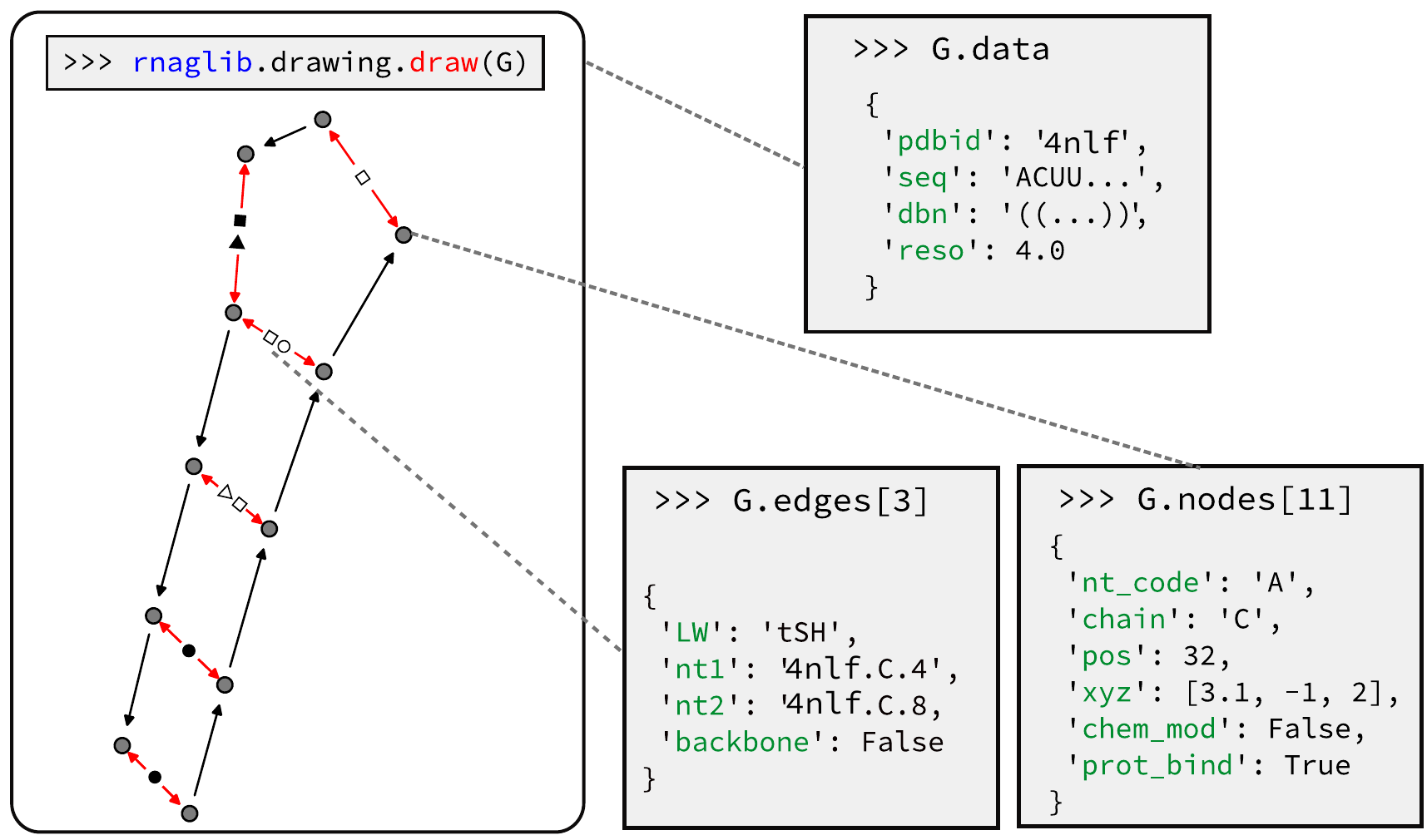}
    \caption{2.5D graph representation of the 23S ribosomal RNA Sarcin Ricin Loop (\texttt{PDBID: 4NLF}). Left panel shows the output of the drawing tool with default settings. Grey boxes are added later to depict a few of the attributes of the graph object, its nodes and its edges.}
    \label{fig:example}
\end{figure}

We collect and update (on a bi-monthly basis) a set of all PDB structures containing RNA on our server.
We include a redundancy reduced subset of graphs based on a list of integrated functional elements curated by the RNA BGSU group (version 3.145; or latest) ~\cite{petrov2013automated}.
We then use the annotation software x3DNA-DSSR ~\cite{lu20033dna} to compute annotations for all structures and complement these annotation using \texttt{BioPython} \cite{cock2009biopython}. 
With this collection of annotations we construct annotated directed 2.5D graphs using the \texttt{networkx} module \cite{hagberg2008exploring}. 
It contains information in each node (such as the nucleotide type or chemical modifications), edge (such as the Leontis-Westhof classification \cite{leontis2001geometric}) and also at the level of the whole graph, such as resolution.
\textbf{Figure \ref{fig:example}} shows an example of such a graph along with the different attributes it contains at the different levels.
This data can be downloaded directly or through our python package.

Equipped with this data, the user can specify which node or edge features need to be included in the graph representation. 
The user can also choose specific target for the machine learning algorithm, and we provide an automatic data splitting routine to test the trained models.
Finally, pre-computations enabling fast subgraphs comparisons are run and also available for use for instance in unsupervised learning settings. 

\section{Machine Learning}
In recent years, graph machine learning tasks are predominantly using graph neural networks. We build our tools into the two most established frameworks for this purpose : PyTorch\cite{paszke2019pytorch} and DGL\cite{wang2019deep}.
The first standardized pipeline we offer is the use of kernel functions for unsupervised machine learning \cite{hamilton2017inductive}.
Unsupervised machine learning settings rely on functions which can compare data points and circumvent the need for annotated data during training.
We have implemented several structural kernels, such as subgraphs matching kernels, along with dedicated data loaders to conduct the data processing specifically for this unsupervised task.

\glib then enables the users to easily combine supervised learning and unsupervised phase, by adding a classification head on the top of the model.
Both parts of this training scheme can be conducted simultaneously, with a loss for each component.

\section{Utility functions}

Along with these machine learning features, we also include several functions to facilitate the handling of RNA 2.5D graphs.
For instance, \glib includes scripts to trim dangling nucleotides, perform statistics on the data or cut a big graph into smaller coherent ones.
\glib also comes with an RNA Graph Edit Distance, the gold standard of graph comparisons, as well as drawing tools customized for 2.5D RNA graphs.

Because \glib comes with a principled way of loading and learning on graphs, we anticipate it can become a reference benchmark for RNA bioinformatics, but also for graph machine learning practitioners.
We include a first baseline on predicting several node level properties and present our results in Table \ref{table:bench2}.
We will keep a leader board of the current best solutions for each of the tasks along with the method used to solve it.

\begin{table}[H]
\centering
\begin{tabular}{lrrrr}
\toprule
\textbf{Task} : & Protein Binding & Small-molecule Binding & Chemical Modification & Link Prediction \\
AuROC & 0.63 & 0.60 & 0.75 & 0.93 \\
\bottomrule
\end{tabular}
\caption{Performance of a baseline model on several tasks of the benchmark. These should be considered as a starting point by practitioners}
\label{table:bench2}
\end{table}

\section{Conclusion}
We present \glib, a set of tools to manipulate graph representations of RNA 3D structures, and use it to conduct machine learning and visualization tasks.
We provide the graph machine learning community with a novel and challenging data set to develop and benchmark new methodologies. 
Not only is it solving a real world problem, it is also a data set whose core signal lies in the graph topology and the edge types, an original setting compared to current graph data sets. 
Simultaneously, we provide the structural RNA community an interface to use graphs and machine learning, hopefully helping this community to better solve RNA challenges.

\section*{Funding}
V. M. is recipient of a doctoral fellowship from the the INCEPTION project [PIA/ANR-16-CONV-0005] and benefits from support from the CRI through "Ecole Doctorale FIRE – Programme Bettencourt". J.W. is supported by a NSERC Discovery grant, FRQ-NT Team grant (PR-284708) and Genome Qu\'ebec Integration Genomics grant.
The authors declare no conflict of interest.

\bibliographystyle{plain}
\bibliography{main}

\begin{thebibliography}{10}

\bibitem{cock2009biopython}
Peter~JA Cock, Tiago Antao, Jeffrey~T Chang, Brad~A Chapman, Cymon~J Cox,
  Andrew Dalke, Iddo Friedberg, Thomas Hamelryck, Frank Kauff, Bartek
  Wilczynski, et~al.
\newblock Biopython: freely available python tools for computational molecular
  biology and bioinformatics.
\newblock {\em Bioinformatics}, 25(11):1422--1423, 2009.

\bibitem{crooke2018rna}
Stanley~T Crooke, Joseph~L Witztum, C~Frank Bennett, and Brenda~F Baker.
\newblock Rna-targeted therapeutics.
\newblock {\em Cell metabolism}, 27(4):714--739, 2018.

\bibitem{djelloul2008automated}
Mahassine Djelloul and Alain Denise.
\newblock Automated motif extraction and classification in rna tertiary
  structures.
\newblock {\em RNA}, 14(12):2489--2497, 2008.

\bibitem{fire1998potent}
Andrew Fire, SiQun Xu, Mary~K Montgomery, Steven~A Kostas, Samuel~E Driver, and
  Craig~C Mello.
\newblock Potent and specific genetic interference by double-stranded rna in
  caenorhabditis elegans.
\newblock {\em nature}, 391(6669):806--811, 1998.

\bibitem{fuller2020amplifying}
Deborah~H Fuller and Peter Berglund.
\newblock Amplifying rna vaccine development.
\newblock {\em New England Journal of Medicine}, 382(25):2469--2471, 2020.

\bibitem{vitravene2002randomized}
Vitravene~Study Group et~al.
\newblock A randomized controlled clinical trial of intravitreous fomivirsen
  for treatment of newly diagnosed peripheral cytomegalovirus retinitis in
  patients with aids.
\newblock {\em American journal of ophthalmology}, 133(4):467--474, 2002.

\bibitem{hagberg2008exploring}
Aric Hagberg, Pieter Swart, and Daniel S~Chult.
\newblock Exploring network structure, dynamics, and function using networkx.
\newblock Technical report, Los Alamos National Lab.(LANL), Los Alamos, NM
  (United States), 2008.

\bibitem{hamilton2017inductive}
William~L Hamilton, Rex Ying, and Jure Leskovec.
\newblock Inductive representation learning on large graphs.
\newblock In {\em Proceedings of the 31st International Conference on Neural
  Information Processing Systems}, pages 1025--1035, 2017.

\bibitem{islam2021rnamotifcontrast}
Shahidul Islam, Md~Mahfuzur Rahaman, and Shaojie Zhang.
\newblock Rnamotifcontrast: a method to discover and visualize rna structural
  motif subfamilies.
\newblock {\em Nucleic Acids Research}, 49(11):e61--e61, 2021.

\bibitem{jumper2021highly}
John Jumper, Richard Evans, Alexander Pritzel, Tim Green, Michael Figurnov,
  Olaf Ronneberger, Kathryn Tunyasuvunakool, Russ Bates, Augustin
  {\v{Z}}{\'\i}dek, Anna Potapenko, et~al.
\newblock Highly accurate protein structure prediction with alphafold.
\newblock {\em Nature}, 596(7873):583--589, 2021.

\bibitem{leontis2001geometric}
Neocles~B Leontis and Eric Westhof.
\newblock Geometric nomenclature and classification of rna base pairs.
\newblock {\em Rna}, 7(4):499--512, 2001.

\bibitem{lu20033dna}
Xiang-Jun Lu and Wilma~K Olson.
\newblock 3dna: a software package for the analysis, rebuilding and
  visualization of three-dimensional nucleic acid structures.
\newblock {\em Nucleic acids research}, 31(17):5108--5121, 2003.

\bibitem{mattick2006non}
John~S Mattick and Igor~V Makunin.
\newblock Non-coding rna.
\newblock {\em Human molecular genetics}, 15(suppl\_1):R17--R29, 2006.

\bibitem{oliver2020augmented}
Carlos Oliver, Vincent Mallet, Roman~Sarrazin Gendron, Vladimir Reinharz,
  William~L Hamilton, Nicolas Moitessier, and J{\'e}r{\^o}me Waldisp{\"u}hl.
\newblock Augmented base pairing networks encode rna-small molecule binding
  preferences.
\newblock {\em Nucleic acids research}, 48(14):7690--7699, 2020.

\bibitem{oliver2020vernal}
Carlos Oliver, Vincent Mallet, Pericles Philippopoulos, William~L Hamilton, and
  Jerome Waldispuhl.
\newblock Vernal: A tool for mining fuzzy network motifs in rna.
\newblock {\em arXiv preprint}, 2020.

\bibitem{paszke2019pytorch}
Adam Paszke, Sam Gross, Francisco Massa, Adam Lerer, James Bradbury, Gregory
  Chanan, Trevor Killeen, Zeming Lin, Natalia Gimelshein, Luca Antiga, et~al.
\newblock Pytorch: An imperative style, high-performance deep learning library.
\newblock {\em Advances in neural information processing systems},
  32:8026--8037, 2019.

\bibitem{petrov2013automated}
Anton~I Petrov, Craig~L Zirbel, and Neocles~B Leontis.
\newblock Automated classification of rna 3d motifs and the rna 3d motif atlas.
\newblock {\em Rna}, 19(10):1327--1340, 2013.

\bibitem{reinharz2018mining}
Vladimir Reinharz, Antoine Soul{\'e}, Eric Westhof, J{\'e}r{\^o}me
  Waldisp{\"u}hl, and Alain Denise.
\newblock Mining for recurrent long-range interactions in rna structures
  reveals embedded hierarchies in network families.
\newblock {\em Nucleic acids research}, 46(8):3841--3851, 2018.

\bibitem{bayespairing}
Roman Sarrazin-Gendron, Vladimir Reinharz, Carlos~G Oliver, Nicolas Moitessier,
  and J{\'e}r{\^o}me Waldisp{\"u}hl.
\newblock Automated, customizable and efficient identification of 3d base pair
  modules with bayespairing.
\newblock {\em Nucleic acids research}, 47(7):3321--3332, 2019.

\bibitem{sarver2008fr3d}
Michael Sarver, Craig~L Zirbel, Jesse Stombaugh, Ali Mokdad, and Neocles~B
  Leontis.
\newblock Fr3d: finding local and composite recurrent structural motifs in rna
  3d structures.
\newblock {\em Journal of mathematical biology}, 56(1):215--252, 2008.

\bibitem{wang2019deep}
Minjie Wang, Lingfan Yu, Da~Zheng, Quan Gan, Yu~Gai, Zihao Ye, Mufei Li,
  Jinjing Zhou, Qi~Huang, Chao Ma, et~al.
\newblock Deep graph library: Towards efficient and scalable deep learning on
  graphs.
\newblock {\em ArXiV}, 2019.

\bibitem{yu2020rna}
Ai-Ming Yu, Young~Hee Choi, and Mei-Juan Tu.
\newblock Rna drugs and rna targets for small molecules: Principles, progress,
  and challenges.
\newblock {\em Pharmacological Reviews}, 72(4):862--898, 2020.

\end{thebibliography}

\end{document}